# Hole-doped $BaFe_{2-x}Cr_xAs_2$ Crystals: A Case of Non-superconductivity

Athena S. Sefat*, David J. Singh*, Lindsay H. VanBebber+, Michael A. McGuire*, Yurij Mozharivskyj°, Rongying Jin*, Brian C. Sales*, Veerle Keppens+, David Mandrus*

*Materials Science & Technology Division, Oak Ridge National Laboratory, Oak Ridge, TN 37831-6114, USA
+ Department of Materials Science & Engineering, University of Tennessee, Knoxville, TN 37996-2200, USA
° Chemistry Department, McMaster University, Hamilton, ON L8S 4M1, Canada

## Abstract

We investigate the physical properties and electronic structure upon Cr-doping in the iron arsenide layers of $BaFe_2As_2$. This form of hole-doping leads to suppression of the magnetic/structural phase transition in $BaFe_{2-x}Cr_xAs_2$ for $x > 0$, but does not lead to superconductivity. For various x values, temperature dependence of the resistivity, specific heat, magnetic susceptibility, Hall coefficient, and single crystal x-ray diffraction data are presented. The materials show signatures of approaching a ferromagnetic state with x, including a metamagnetic transition for x as little as 0.36, an enhanced magnetic susceptibility, and a large Sommerfeld coefficient. Such results reflect renormalization due to spin fluctuations and they are supported by density functional calculations at $x = 1$. Calculations show a strong interplay between magnetic ordering and chemical ordering of Fe and Cr, with a ferromagnetic ground state. This ferromagnetic ground state is explained in terms of the electronic structure. The resulting phase diagram is suggestive that superconductivity does not derive simply from the suppression of the structural/magnetic transitions.

## 1. Introduction

The discovery of high-$T_C$ superconductivity in the Fe-based systems has attracted great interest primarily due to their non copper-based origin. The structural parents of such superconductors include those of tetragonal $ThCr_2Si_2$-type $AFe_2As_2$ [1], the so-called '122' family. The stoichiometric compounds exhibit a coupled structural transition with an antiferromagnetic spin-density-wave ordering of Fe-spins at 171 K in $A$ = Ca [2], 205 K in $A$ = Sr [3], 132 K or 140 K in $A$ = Ba [4, 5], and at 200 K in $A$ = Eu [6]. The onset of superconductivity in 122 family with suppression of the magnetic ordered state is similar to the behavior observed in the cuprates, and can be induced by applying pressure [7], or hole-doping on the $A$ site [8], or electron-doping on Fe site [9]. The appearance of superconductivity is associated with the suppression of the structural/magnetic phase transitions. The induced superconductivity due to in-plane cobalt doping provides one crucial distinction from the cuprates [9]. In addition, the itinerant and weakly correlated electrons in $AFe_2As_2$ parents distinguish them from the cuprate Mott insulating parents.

In this paper we investigate the properties of $Ba(Fe,Cr)_2As_2$ solid solution by hole-doping in the FeAs layers of $BaFe_2As_2$. This system allows the exploration of the structural/magnetic phase transitions, and its possible relation to superconductivity. Compared to $Fe^{2+}$ ion ($3d^6$), $Cr^{2+}$ has two less $3d$ electrons. Thus it is expected that Cr doping introduces disorder in the iron arsenide layer, and changes both crystal and electronic structures. The crystallographic and electronic structure, thermodynamic and transport properties of Cr-doped $BaFe_2As_2$ crystals are presented.

While for undoped $BaFe_2As_2$ a single structural/magnetic transition is observed at $T_N$ = 132(1) K in specific heat [4, 5], for x = 0.04, the transition is shifted to 116 K. Much to our surprise, single crystal x-ray diffraction data at 100 K gives no detectable symmetry-breaking transition for this composition. With more Cr substitution in $BaFe_{2-x}Cr_xAs_2$, the magnetic transition temperature continues to decrease. For x = 0.75, there is no evidence of long-range magnetic order, nor superconductivity down to 1.8 K. Theory suggests that $BaCr_2As_2$ is an itinerant antiferromagnetic metal [10]. The stronger Cr-As covalency relative to $BaFe_2As_2$ may explain why superconductivity is not observed in Cr-doped $BaFe_2As_2$. Large single crystals of $BaFe_{2-x}Cr_xAs_2$ with $0 \leq x \leq 0.75$ are grown to study the systematic property changes.

## 2. Results and Discussion

The single crystals of $BaFe_{2-x}Cr_xAs_2$ were grown out of a mixture of FeAs and CrAs flux. High purity elements (>99.9%, from Alfa Aesar) were used in their preparations. The FeAs and CrAs binaries were synthesized similarly to our recent report [9]. To produce a sample with a nominal composition ($x_{nomin}$) of 0.06, a ratio of Ba:FeAs:CrAs = 1:4.85:0.15 was used; for $x_{nomin}$ = 0.12, this ratio was 1:4.7:0.3; for $x_{nomin}$ = 0.22, this ratio was 1:4.45:0.55. Each of these mixtures was heated in an alumina crucible for 13 hours at 1230°C under a partial atmosphere of argon. Each reaction was then cooled at a rate of 1.8°C/hour, followed by decanting the FeAs/CrAs flux at 1100°C. In order to produce a sample with $x_{nomin}$ = 0.14, a ratio of Ba:FeAs:CrAs = 1:4.3:0.7 was used; for x = 0.28, this ratio was 1:3.6:1.4; for $x_{nomin}$ = 0.5, this ratio was 1:3.75:1.25; for $x_{nomin}$ = 1, this ratio was 1:2.5:2.5. Each of these mixtures was heated for 10 hours at 1230°C and then cooled at a rate of 2°C/hour, followed by decanting of the flux at 1120°C. The crystals had sheet morphologies and dimensions of ~ 6 × 5 × 0.15 $mm^3$ or smaller in *a, b*, *c* directions, respectively. Similar to $BaFe_2As_2$ [9], the crystals formed with thin face perpendicular to *c*-axis. Attempts of crystal growth for $x_{nominal}$ > 1.0 values were unsuccessful, but may be possible by revisiting the synthetic routes.

The chemical composition was measured with a JEOL JSM-840 scanning electron microscope, same as that described in Ref. [9]. Energy-dispersive x-ray spectroscopy (EDS) analyses indicated that less Cr was substituted in the crystal than put in solution; these results are summarized in Table 1. The samples will be denoted by these measured values of x throughout this manuscript.

The phase purity of the crystals was characterized using a Scintag XDS 2000 powder x-ray diffractometer. Lattice constants were determined from full-pattern LeBail refinements using the program FullProf [11]. At room temperature the structures are identified as the tetragonal $ThCr_2Si_2$ (*I*4/*mmm*, Z = 2). Table 1 gives the refined lattice constants from powder x-ray diffraction; Fig. 1 plots *a*- and *c*- parameters as a function of Cr concentration. The incorporation of x = 0.75 Cr in the Fe site increases the cell volume by 2.6%, mainly due to the increase of the *c* lattice constant. Single crystal x-ray diffraction was also done at room temperature for x = 0.75. Analysis of the diffraction spots of the crystals indicated that there were multiple domains (plates) with preferred stacking giving some smeared reflections. The refined lattice parameters are $a$ = 3.985(3), and $c$ = 13.277(17) Å, similar to powder x-ray diffraction results (Table 1). Due to similarities between the Fe and Cr atomic scattering powers, the Fe/Cr ratios could not be refined.

Low temperature single crystal x-ray diffraction data was collected on the parent $BaFe_2As_2$. The crystal was placed under a stream of nitrogen gas and the data were collected on a Bruker Apex II diffractometer, first at 100 K. The refinement of the reflections gives the orthorhombic *F*-centered cell (*Fmmm*), as reported [5]. The refined lattice parameters are $a$ = 5.595(3) Å, $b$ = 5.610(3) Å, and $c$ = 12.964(6) Å. The crystal was then warmed to 173 K; this structure was found to be tetragonal (*I*4/*mmm*) with lattice parameters of $a$ = 3.9708(16) Å, and $c$ = 13.006(5) Å, as expected.

For small Cr-doping of x = 0.04, low temperature structure was also examined by single crystal x-ray diffraction. A crystal was first examined at 100 K, then by warming to 117 K, 119 K, and finally 173 K. Surprisingly the 100 K structure refined in the tetragonal *I*4/*mmm*, with no sign of distortion. Shapes of several diffraction spots from 100 K and 173 K were compared and they are almost identical. Also, the thermal ellipsoids are quite similar for 100 K and 173 K in the *I*4/*mmm*. A small amount of chromium dopant seems to suppress a symmetry-breaking structural transition in $BaFe_2As_2$. If there is any distortion to a lower symmetry, it is relatively small and undetectable from the single crystal data. Fig. 2 shows the lattice constants as a function of temperature for x = 0.04. There is an indication that some structural change may happen below 119 K since both cell parameters decrease. High-resolution powder diffraction may reveal peak splitting (orthorhombic or another distortion) if present. If a symmetry lowering does not occur, the concomitant decrease in both $a$ and $c$ suggest that the structural transition may be volumetric in nature.

Magnetic measurements were performed with a Quantum Design (Magnetic Property Measurement System) SQUID magnetometer. For a temperature sweep experiment, the sample was cooled to 1.8 K in zero-field (zfc) and data were collected by warming from 1.8 K to 300 K in an applied field of 1 Tesla. The magnetic susceptibility results are presented per mole of formula unit ($cm^3$/mol), $\chi$, along $c$- and $ab$-crystallographic directions (Fig. 2a, b). For $BaFe_2As_2$ at room temperature $\chi_c \approx \chi_{ab} \approx 6.5 \times 10^{-4}$ $cm^3$ $mol^{-1}$, and the susceptibility decreases linearly with decreasing temperature, and drops abruptly below $T_N \approx 132$ K with $\chi_c > \chi_{ab}$. With Cr-doping, the magnitude of the magnetic susceptibility increases at room temperature while the transition for magnetic order (probably $T_N$) decreases. At 1.8 K and for $x \leq 0.14$, $\chi_c$ is approximately 1.2 times larger than $\chi_{ab}$. With more Cr-doping, the susceptibility becomes more anisotropic: for x = 0.20 the ratio of $\chi_c/\chi_a = 1.4$ (at 1.8 K), and for x = 0.36 it is $\chi_c/\chi_{ab} = 1.8$ (at 1.8 K). The susceptibilities along both crystallographic axes roughly converge and are isotropic above $T_N = 118$ K for x = 0.04, $T_N = 104$ K for x = 0.08, $T_N = 88$ K for x = 0.14, $T_N = 80$ K for x = 0.20, and $T_N = 58$ K for x = 0.36. For x = 0.75 and at ~ 120 K there is a drop in $\chi_c$ and a small kink in $\chi_{ab}$. For this sample there is also a broad feature at ~ 30 K in $\chi_{ab}$. For x = 0.75, the magnetic anisotropy switches, with $\chi_{ab}/\chi_c \approx 2$ at 1.8 K. Figs. 2b and 2c illustrate the field dependent magnetization at 1.8 K. For samples with $x \leq 0.20$, the behavior is linear. For x = 0.36 ($M_{ab}$) and x = 0.75 ($M_c$), there are non-linearities in magnetization indicating the possibility of metamagnetic transitions. These were not pursued any further in this work.

Transport measurements were performed with a Quantum Design Physical Property Measurement System (PPMS). Electrical leads were attached to the sample using Dupont 4929 silver paste and resistivity measured in the *ab* plane (Fig. 3a). The $\rho_{300K}$ values range from 0.3 mΩ cm to 0.8 mΩ cm, although their absolute values may suffer from the geometry factor estimations. Temperature dependent electrical resistivity for $BaFe_2As_2$ decreases with decreasing temperature, and below ~ 132 K there is a sharp drop. For x > 0, instead of a drop in $\rho_{ab}$, there are sharp upturns below 117 K for x = 0.04, 104 K for x = 0.08, 90 K for x = 0.14, and 60 K for x = 0.36. The temperature of the anomaly is clearly diminished with increasing x, with no evidence of superconductivity down to 1.8 K. The resistivity for x = 0.75 sample is roughly temperature independent down to ~ 100 K, and rises below. The difference between the resistivity anomaly for x = 0 and those of $0.04 \leq x \leq 0.36$ is probably associated with changes in the scattering and the decrease in the number of carriers.

Fig. 3b illustrates the temperature dependence of the Hall coefficient. The Hall voltage was calculated from the anti-symmetric part of the transverse voltage (perpendicular to the applied current) under magnetic-field (± 6 Tesla) reversal at fixed temperature. For x = 0, negative Hall coefficient is reported in temperature region of 125-145 K, less negative above 134 K [13]. For x = 0.04, $R_H$ is negative between 30 and 300 K. $R_H$ becomes more negative with decrease of temperature with a sharp anomaly at ~ 120 K, it then gives a minimum near 90 K, becoming less negative thereon. For x = 0.08, x = 0.14, and x = 0.36, the $R_H$ values become more positive below 104 K, 90 K, and 60 K, respectively, and are approximately temperature independent above. These values coincide with $T_N$ observed in $\chi(T)$ and anomalies observed in $\rho(T)$. For x = 0.75, $R_H$ is roughly temperature independent and positive down to 5 K. Further interpretation of $R_H$ is complicated by the multi-band nature of the system and the likely presence of both electron and hole bands at the Fermi level.

Specific heat data were collected on single crystals (Fig. 4a), also using a PPMS. For $BaFe_2As_2$, a second-order-like transition is observed at 132(1) K, associated with a tetragonal to orthorhombic *Fmmm* distortion, and a spin density wave (SDW) magnetic transition [4, 12]. With Cr-doping, the transition temperatures decrease. For x = 0.75, there are no anomalies in specific heat down to 1.8 K (Fig. 4a, inset). Fig. 4b plots the C/T versus $T^2$ dependence. The fitted Sommerfeld coefficient, $\gamma$, for the linear region below 6 K gives 6.1 mJ/($K^2$mol) [or 3.0 mJ/($K^2$mol Fe)] for $BaFe_2As_2$ [12]. The $\gamma$ value increases consistently with Cr-doping: $\gamma$ = 30.1 mJ/($K^2$mol) for x = 0.14, $\gamma$ = 38.3 mJ/($K^2$mol) for x = 0.20, 65.4 mJ/($K^2$mol) for x = 0.36, and largest at 69.9 mJ/($K^2$mol) [or 35.0 mJ/($K^2$mol transition metal)] for x = 0.75.

Having both $\gamma$ and magnetic susceptibility data, we may further estimate the Wilson ratio $R_w = \pi^2 k_B^2 \chi/(3\mu_B^2\gamma)$, assuming that $\chi_{spin} \approx \chi$ measured at 1.8 K. For x = 0.14, $R_w$ is ~ 5 from $\chi_{ab}$ and ~ 6 from $\chi_c$. For x = 0.20 $R_w$ is ~ 6 from $\chi_{ab}$ and ~ 8 from $\chi_c$. For x = 0.36 $R_w$ is ~ 9 from $\chi_{ab}$ and ~ 16 from $\chi_c$. Finally for x = 0.75 $R_w$ is ~ 13 from $\chi_{ab}$ and ~ 7 from $\chi_c$. These values significantly exceed unity for a free electron system and indicate that $BaFe_2As_2$ approaches ferromagnetism upon Cr-doping. We note that the nearness to ferromagnetism and ferromagnetic spin fluctuations are highly destructive to singlet superconductivity, including s, s+/- and d wave states.

For $BaFe_2As_2$ there is a single sharp peak in $C_p$ consistent with $d(\chi T)/dT$ and $\rho(T)$ results (Fig. 5a). With a small amount of Cr-doping, this single sharp peak gives two distinct features in $C_p$ for x = 0.04 (Fig. 5b): a sharper peak at 115.8 K followed by a shoulder at 118.3 K. From the overlapped data of $\rho(T)$ and $d(\chi T)/dT$, features are observed roughly at the lower transition temperature, and we assign $T_N \approx 116$ K. As discussed earlier, no detectable symmetry breaking transition occurs for this composition in single crystal x-ray diffraction.

Therefore the peak splitting effect in specific heat may be associated with phase separation due to a variation in Cr concentration within the crystal. Amongst the specific heat results, x = 0.20 is the only other composition that gives the double peak feature (Fig. 5c), with $T_N \approx 78$ K and a shoulder at 79 K. For x = 0.14, there is a peak at $T_N \approx 87.5$ K. For x = 0.36, $T_N \approx 57.3$ K.

Based on the data above, a composition-temperature (x-T) phase diagram is proposed for the hole-doped $BaFe_{2-x}Cr_xAs_2$, shown in Fig. 6. The nature of the magnetic transition temperature is probably of antiferromagnetic SDW type. In order to confirm this, neutron diffraction studies need to be done.

First principles supercell calculations for x = 1 were performed within the local density approximation (LDA) with the general potential linearized augmented planewave method (LAPW), similar to those described previously [10, 14]. The calculations were done using lattice parameters of $a$ = 3.979 Å and $c$ = 13.26 Å. We considered different magnetic orders, specifically, ferromagnetic, *A*-type antiferromagnetic (ferromagnetic layers stacked antiferromagnetically along the *c*-axis), *G*-type antiferromagnetic (nearest neighbor antiferromagnetism in all directions, which is the ground state of $BaMn_2As_2$ [15] and $BaCr_2As_2$ [10]), and a stripe antiferromagnetism, as in the spin density wave ground state of $BaFe_2As_2$. We considered three different chemical ordering patterns – a checkerboard ordering of Fe and Cr, with alternation of Fe and Cr along the *c*-axis, and a stripe ordering (lines of Fe and Cr neighbors), with both possible stackings along *c*. The atomic positions in the unit cell were separately relaxed for each chemical and magnetic order. The lowest energy state is for the checkerboard chemical ordering and ferromagnetism. In contrast, with stripe chemical ordering, the lowest energy state is antiferromagnetic (AFM) with *G*-type ordering. This, however, is 0.10 eV/formula unit higher in energy than the checkerboard chemical ordering. Considering the growth temperature, this indicates that at least partial chemical ordering may be expected.

Turning to the magnetic properties, we find substantial Fe and Cr magnetic moments independent of chemical order. For the (non-ground state) stripe chemical order, we obtained a *G*-type AFM state, with Fe and Cr moments of 1.96 $\mu_B$ and 2.31 $\mu_B$, respectively, as defined by the integrals of the spin densities in the LAPW spheres of radius 2.1 Bohr. This state was 0.045 eV lower in energy than the ferromagnetic solution for that chemical ordering. For the low energy checkerboard chemical order, the ferromagnetic solution is favored over the *G*-type solution by 0.160 eV, which is a very large magnetic energy. Furthermore, the moments are strongly dependent on both chemical and magnetic order. For the lowest energy ferromagnetic state with checkerboard chemical order, we obtain Fe and Cr moments of 0.97 $\mu_B$ and 1.80 $\mu_B$, respectively, while for the *G*-type magnetic order order, the Fe and Cr moments become 1.78 $\mu_B$ and 0.07 $\mu_B$. On the other hand, the energy difference between the calculated lowest energy ferromagnetic state, and the *A*-type AFM order is very small: 0.003 eV/formula unit indicating rather weak *c*-axis interactions, at least for this chemical order. Also considering the small value of this energy difference and the limitations of density functional calculations, we cannot exclude an *A*-type AFM ground state. In any case, these results show that there is a very strong interplay between magnetism and chemical ordering in this material, and, based on sensitivity of the magnetic moments to the order, that the magnetism has a strong itinerant component. Finally, we note that the properties of $BaFeCrAs_2$ (x = 1) are very different from those of $BaMn_2As_2$, even though the electron count is the same. This is in contrast to the behavior seen in alloys with Fe, Co and Ni.

We now discuss the ferromagnetic state of chemically ordered $BaFeCrAs_2$ (x = 1) in more detail. The calculated band structure, density of states (DOS) and Fermi surface are shown in Figs. 7, 8, and 9, respectively. The band structure and Fermi surface show a clearly metallic state, with several bands crossing the Fermi energy, and large Fermi surfaces. Furthermore, in conjunction with the projected DOS, one may note a strong spin dependent hybridization, both with As and between the Fe and Cr *d* orbitals. In particular, one may note that the majority spin DOS shows very similar coherent shapes of the Fe and Cr contributions, while the minority spin does not. This formation of a coherent metallic majority spin band structure generally leads to a reduction in band energy and explains why the ferromagnetic ordering is strongly favored. This basic mechanism is closely related to stabilization mechanism for the ferromagnetic metallic state in La rich alloys of $(La,Ca)MnO_3$ for example [16]. In any case, it also provides and explanation of why there is a strong interplay between chemical and magnetic order. Specifically, with checkerboard chemical order direct nearest neighbor hopping can only take place between Fe and Cr atoms. This strongly favors a ferromagnetic state, as discussed. On the other hand, for stripe-like chemical ordering, the strong covalency between Cr and As, noted previously [10] provides a strong antiferromagnetic interaction between neighboring Cr atoms works against a ferromagnetic state, and so an antiferromagnetic state emerges with strongly enhanced moments.

The majority spin Fermi surface is dominated by two large electron cylinders running along the zone corner and shows additionally small three dimensional electron sections centered at the Z points (0,0,1/2) and some very tiny electron sections. In contrast to the majority spin which is dominated by electron cylinders, the minority spin Fermi surface shows a prominent multiband character, with a hole cylinder around the zone center and a large electron cylinder around the zone corner. In addition, there are three dimensional hole pockets around the Γ point and additional tiny hole sections. The DOS at the Fermi energy, $N(E_F)$ = 3.1 $eV^{-1}$/formula unit of which 29% comes from the majority spin and the remainder from the minority spin. This corresponds to a bare band specific heat $\gamma$ = 7.3 mJ/($K^2$ mole formula unit), i.e. 3.65 mJ/($K^2$ mole transition element). This is much less than measured values for our high Cr content samples, indicating a very strong mass renormalization. The calculated Fermi velocities in the *ab*-plane are 2.3 x $10^5$ m/s and 1.5 x $10^5$ m/s for majority and minority spin, respectively, while the corresponding *c*-axis values are 1.2 x $10^5$ m/s and 0.9 x $10^5$ m/s. These numbers indicate a resistivity anisotropy $\rho_c/\rho_{ab} \sim 3$ to 4 depending on the scattering, which may be spin dependent in this material.

## 3. Conclusion

As chromium is introduced into FeAs layers in $BaFe_{2-x}Cr_xAs_2$, $T_N$ is suppressed while specific heat $\gamma$ increases. The latter is indicative of strong scattering due to a combination of disorder and strong spin fluctuations, which are generally enhanced in cases where there is competition with another magnetic state. The metamagnetic transitions and the enhanced susceptibilities at x = 0.36 and x = 0.75 levels imply that with increasing x, a ferromagnetic state may be approached. This is consistent with the ferromagnetic ground state that we find for the chemically ordered x = 1. Considering the strong interplay between chemical order and itinerant magnetic order, as well as the strong spin dependent hybridization that underlies this, one may expect (1) strong magnetic scattering both because of spin disorder connected with chemical disorder, and also due to spin fluctuations, and (2) the possibility of glassy magnetic behavior in material that is away from the x = 1 chemical composition and also in material with partial chemical order. In this regard, it will be very interesting to explore the dependence of the properties on stoichiometry for higher Cr contents closer to the x = 1 composition and also to perform annealing studies to investigate the interplay between magnetic order, transport and chemical order. The present results clearly show a competition between the SDW-type magnetism and another state that is ferromagnetic or very nearly ferromagnetic as Cr is introduced into $BaFe_2As_2$, as well as strong scattering and specific heat renormalization in the regime where the magnetic cross-over occurs. The theoretical results show that ordered $BaFeCrAs_2$ is a rather unusual itinerant ferromagnetic metal with a strong spin dependent hybridization and an interesting interplay between magnetic and chemical order.

**Acknowledgement**

The research at ORNL was sponsored by the Division of Materials Sciences and Engineering, Office of Basic Energy Sciences, U.S. Department of Energy. Part of this research was performed by Eugene P. Wigner Fellows at ORNL, managed by UT-Battelle, LLC, for the U.S. DOE.

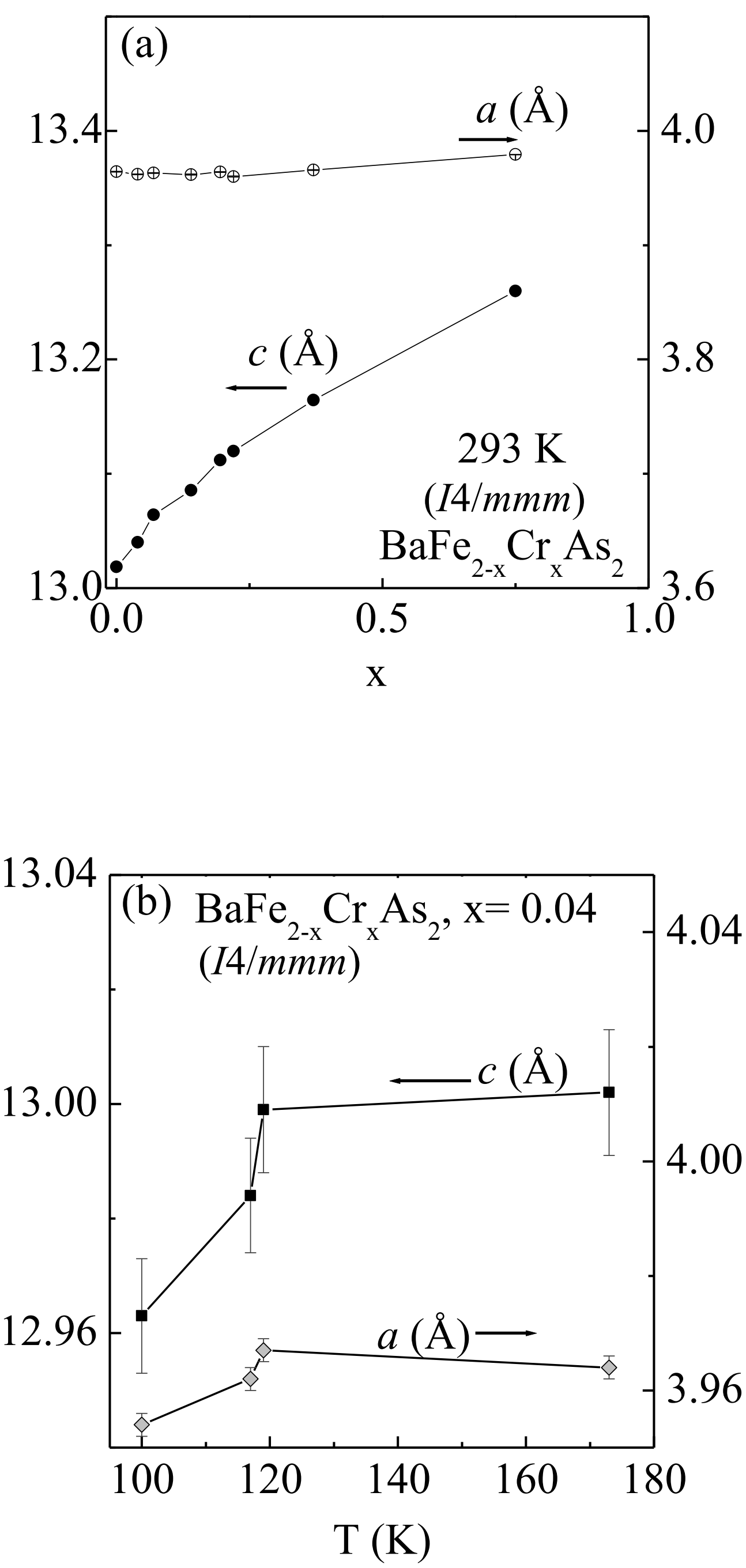

Fig. 1: (a) Lattice parameters for $BaFe_{2-x}Cr_xAs_2$ compositions at room temperature, refined from powder x-ray diffraction data. (b) The change of lattice parameters with temperature for x = 0.04 in $BaFe_{2-x}Cr_xAs_2$, refined in the tetragonal lattice.

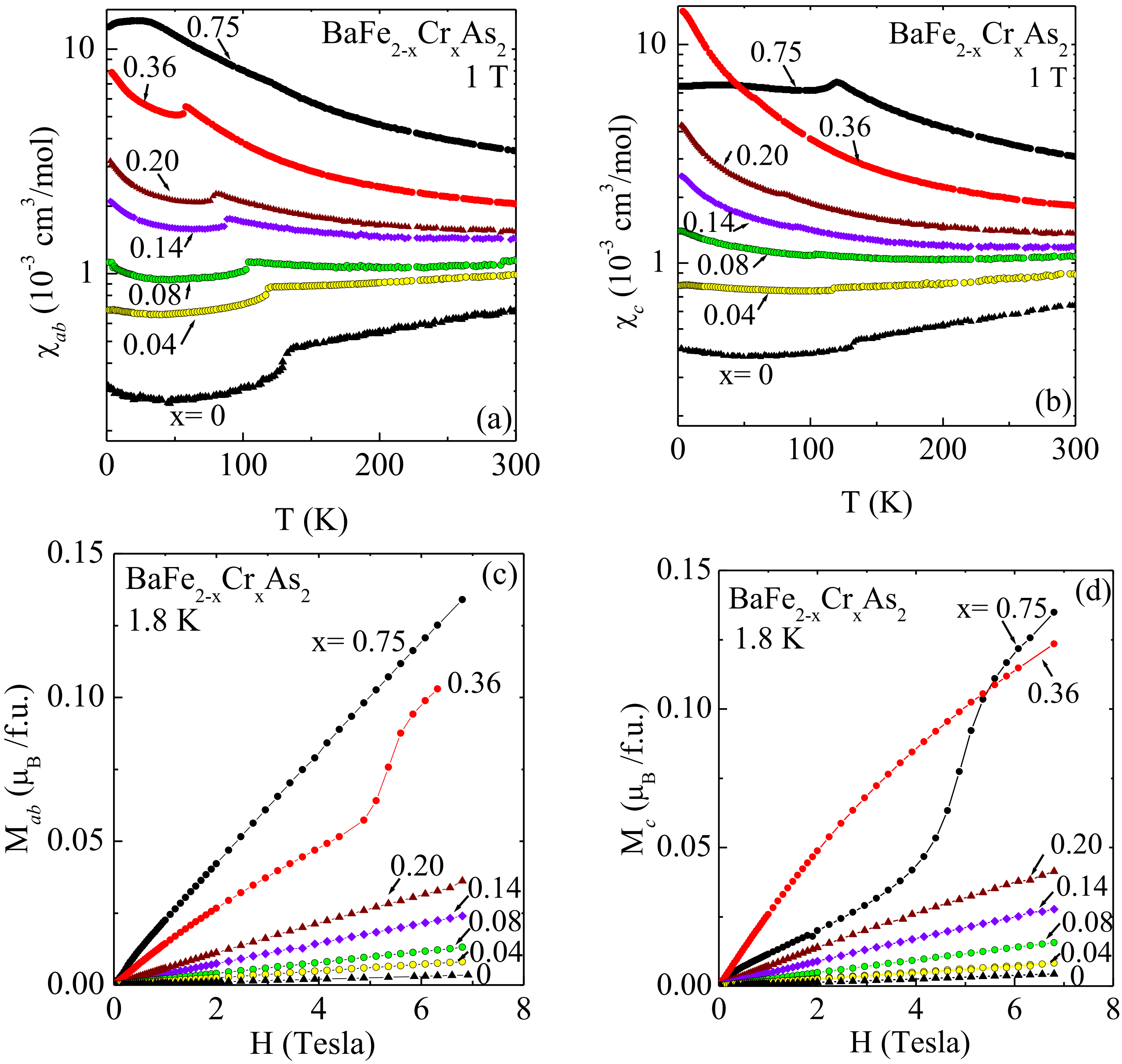

Fig. 2: (Color online) Temperature dependence of molar susceptibility for $BaFe_{2-x}Cr_xAs_2$ between $0 \leq x \leq 0.75$, along *ab*- (a) and *c*-lattice directions (b). Magnetization versus applied field measured at 1.8 K for $BaFe_{2-x}Cr_xAs_2$, along *ab*- (c) and *c*-lattice directions (d).

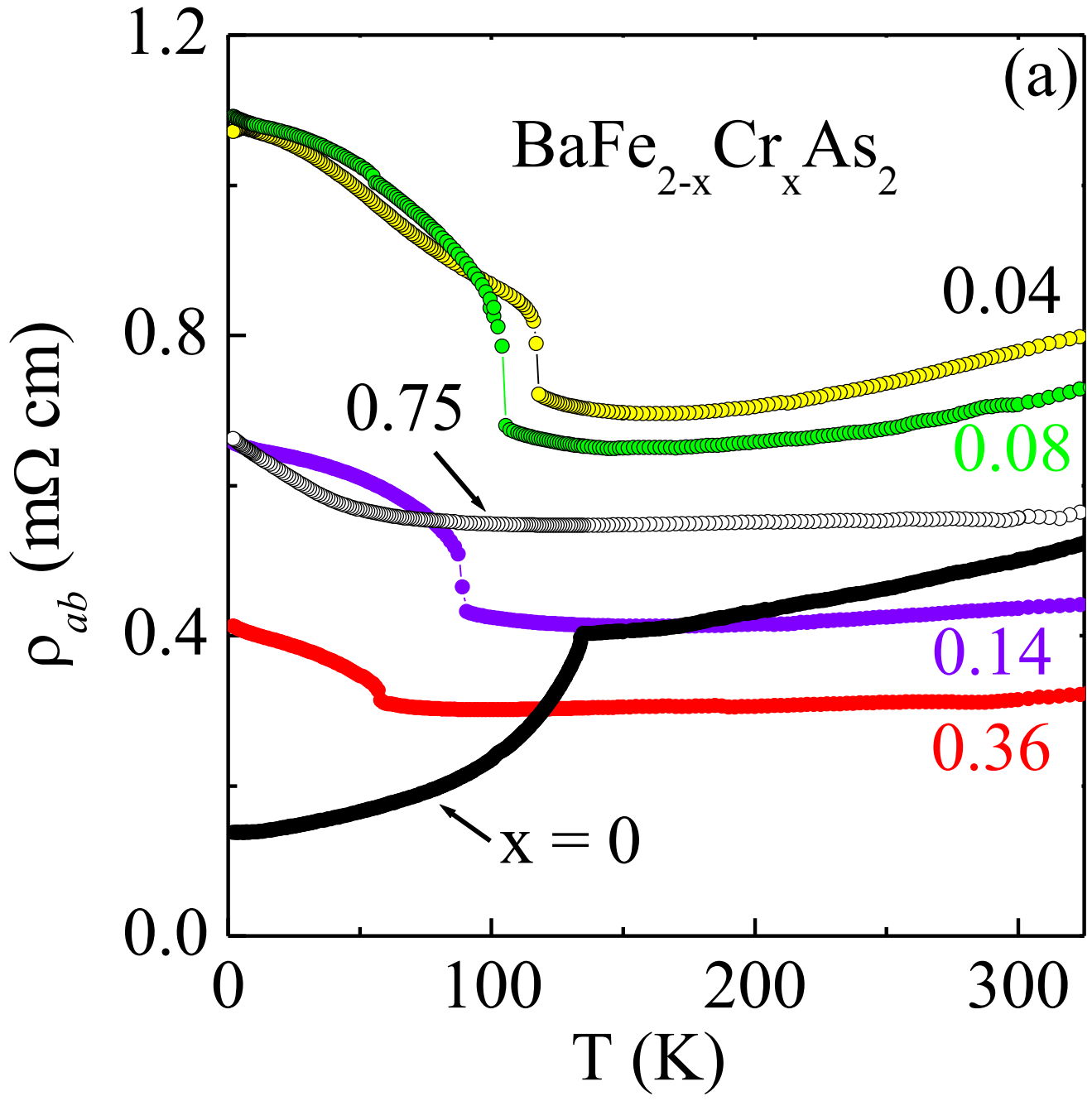

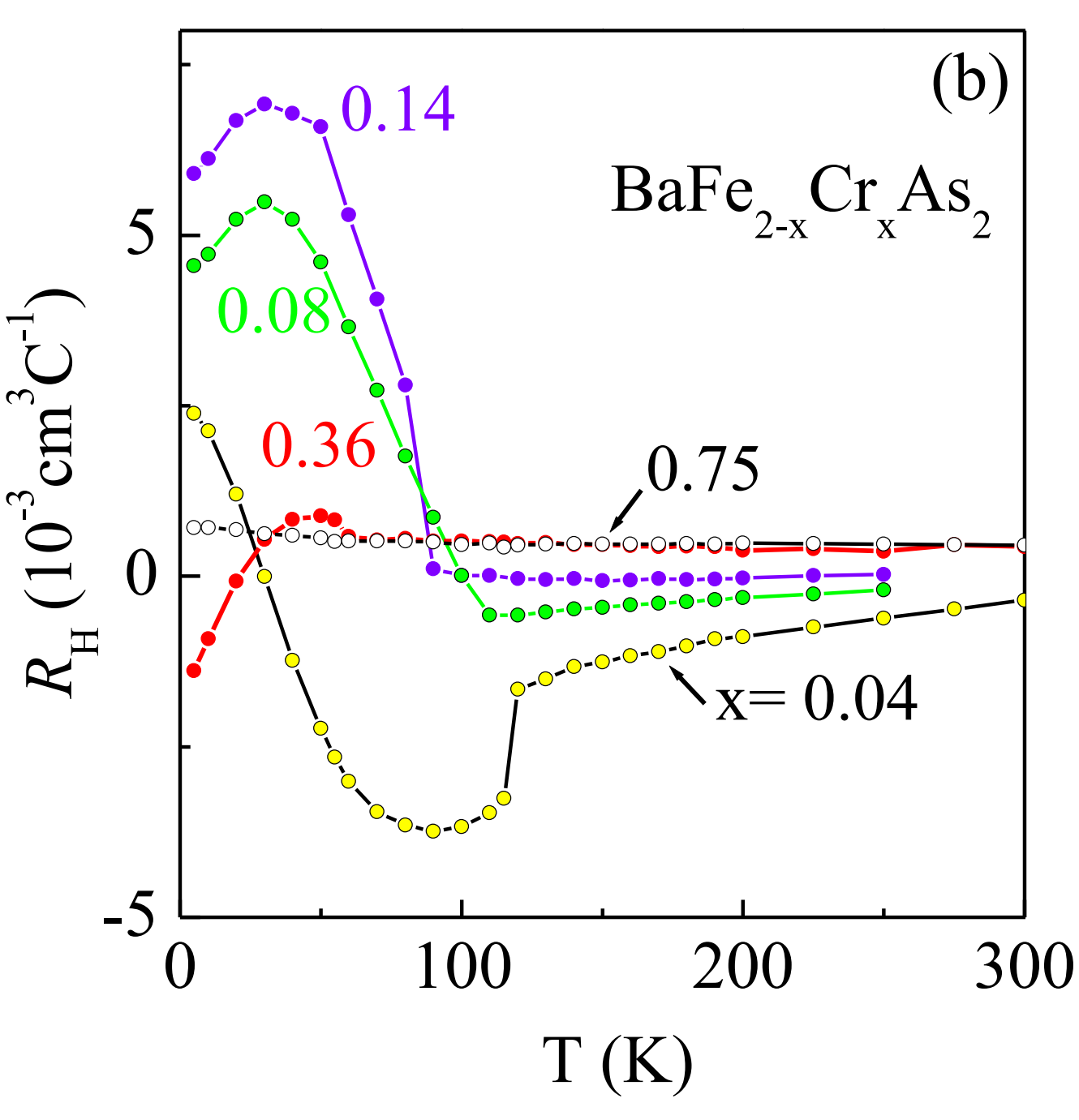

Fig. 3: (Color online) Temperature dependence of (a) resistivity, and (b) Hall coefficient ($R_H$) for $BaFe_{2-x}Cr_xAs_2$ compositions between $0 \leq x \leq 0.75$, measured in the *ab*-plane.

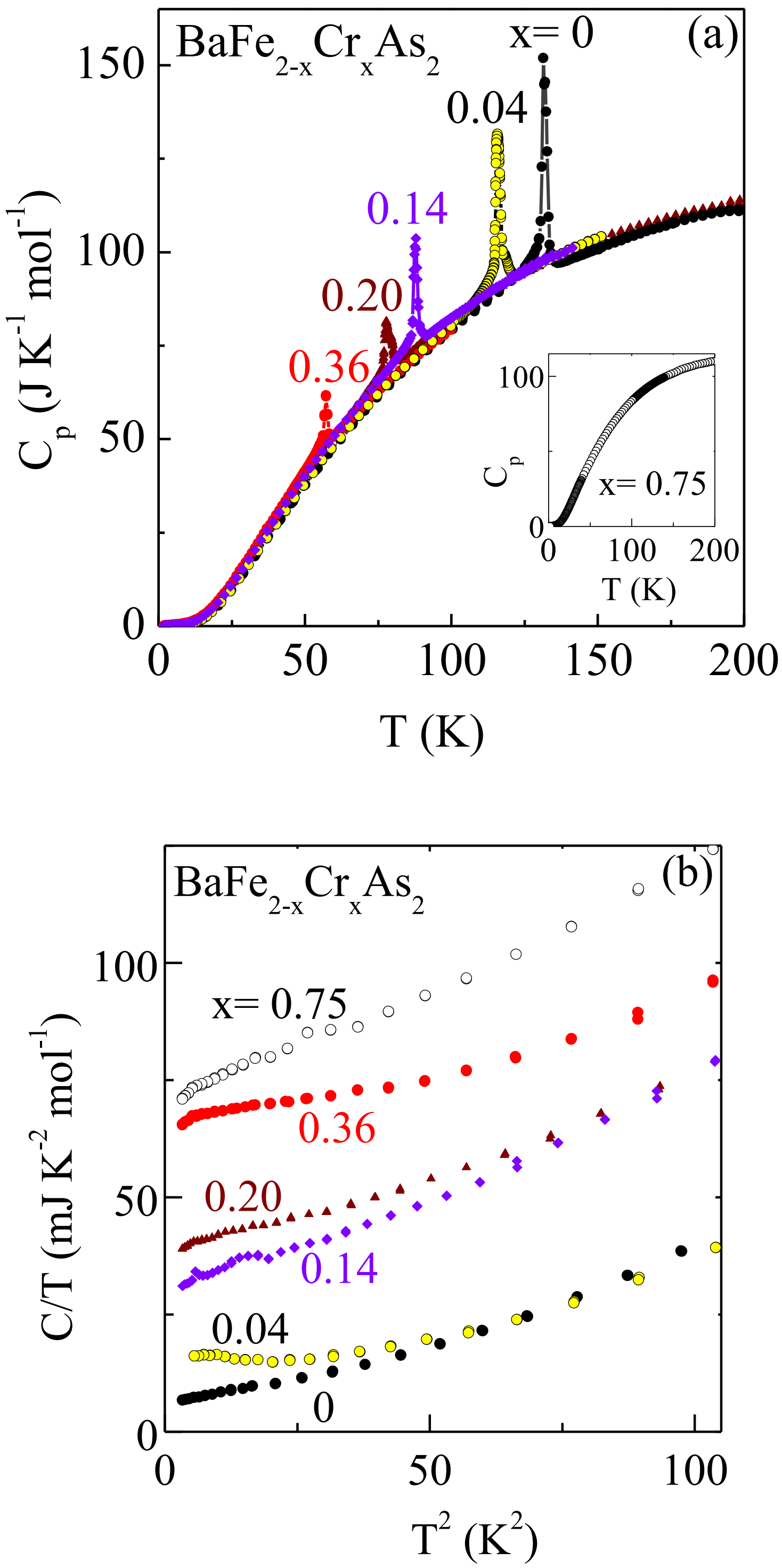

Fig. 4: (Color online) For $BaFe_{2-x}Cr_xAs_2$ and compositions between $0 \le x \le 0.75$, temperature dependence of specific heat in (a) $C_p(T)$ form shown below 200 K, and in (b) $C_p/T$ versus $T^2$ form shown below ~10 K.

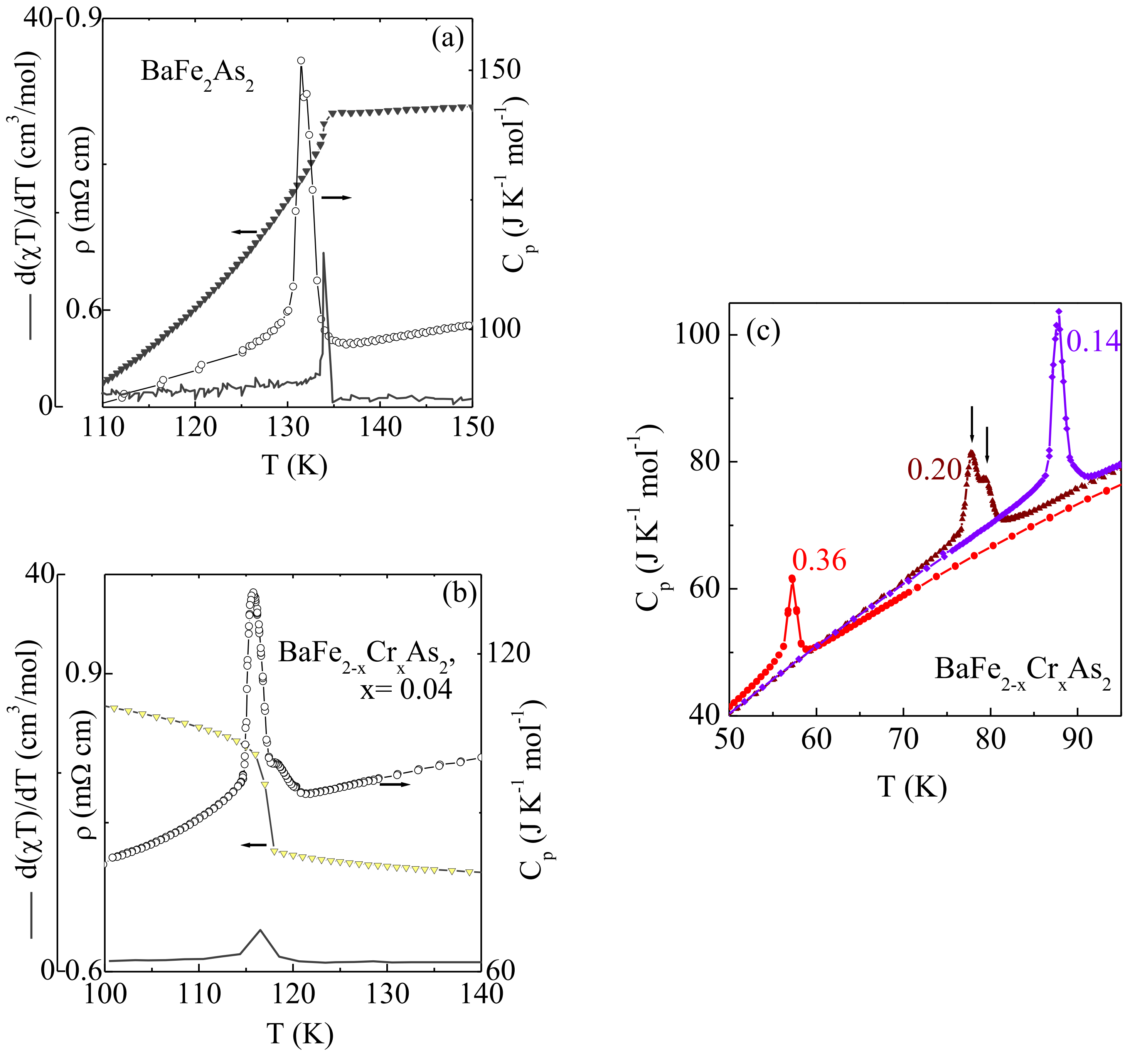

Fig. 5: (Color online) For BaFe$_{2-x}$Cr$_x$As$_2$, the overlapped data of ρ(T), $C_p$(T), and $\chi_{ab}$ shown in the form d(χT)/dT for (a) x = 0 and (b) x = 0.04. (c) Enlarged $C_p$(T) region for x = 0.14, 0.20, and 0.36; the double peak feature is noted by arrows for x = 0.20.

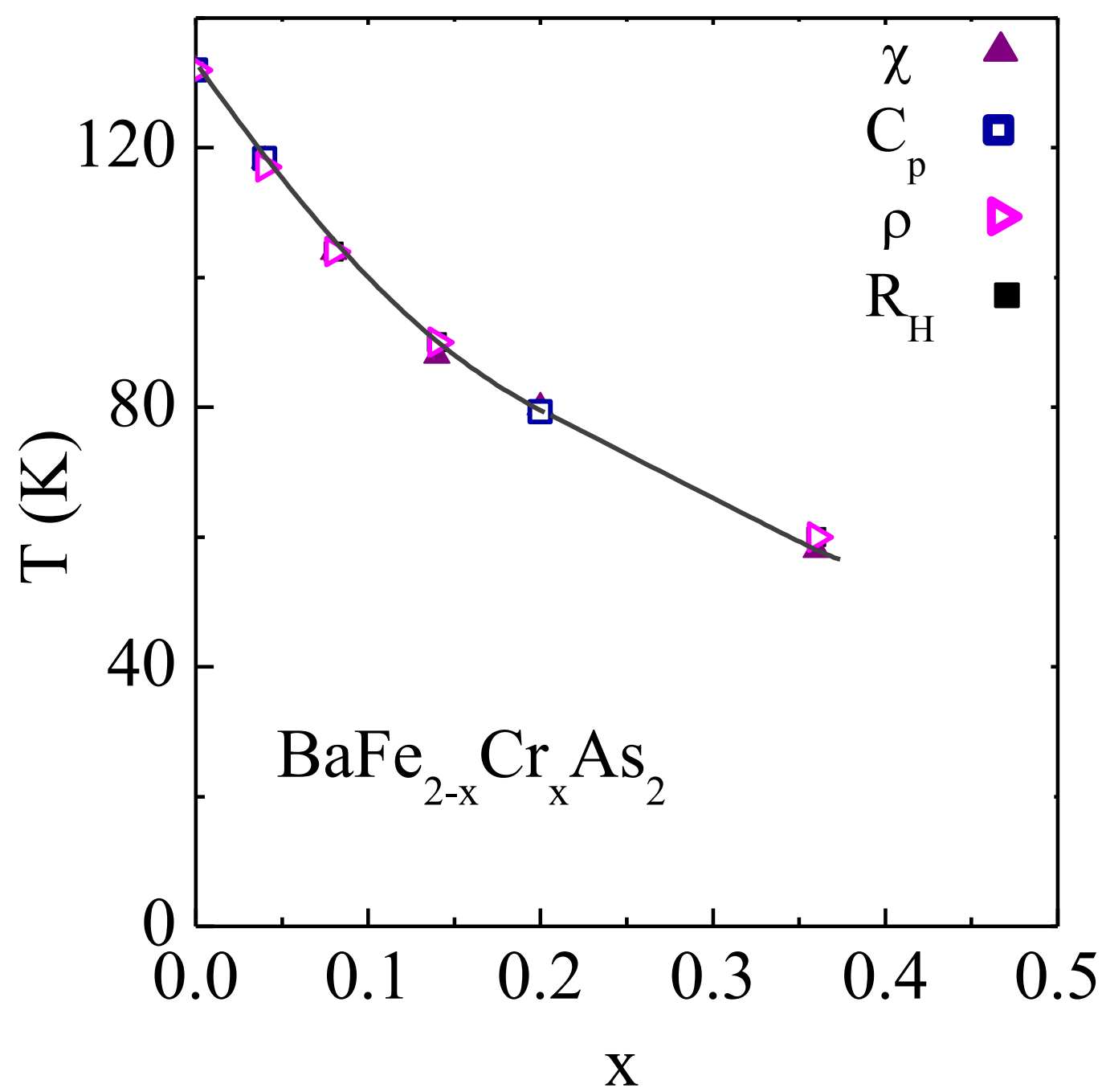

Fig. 6: (Color online) For $BaFe_{2-x}Cr_x As_2$, magnetic ordering temperature ($T_N$) versus Cr-doping (x). For x = 0.75, long-range order seems lost.

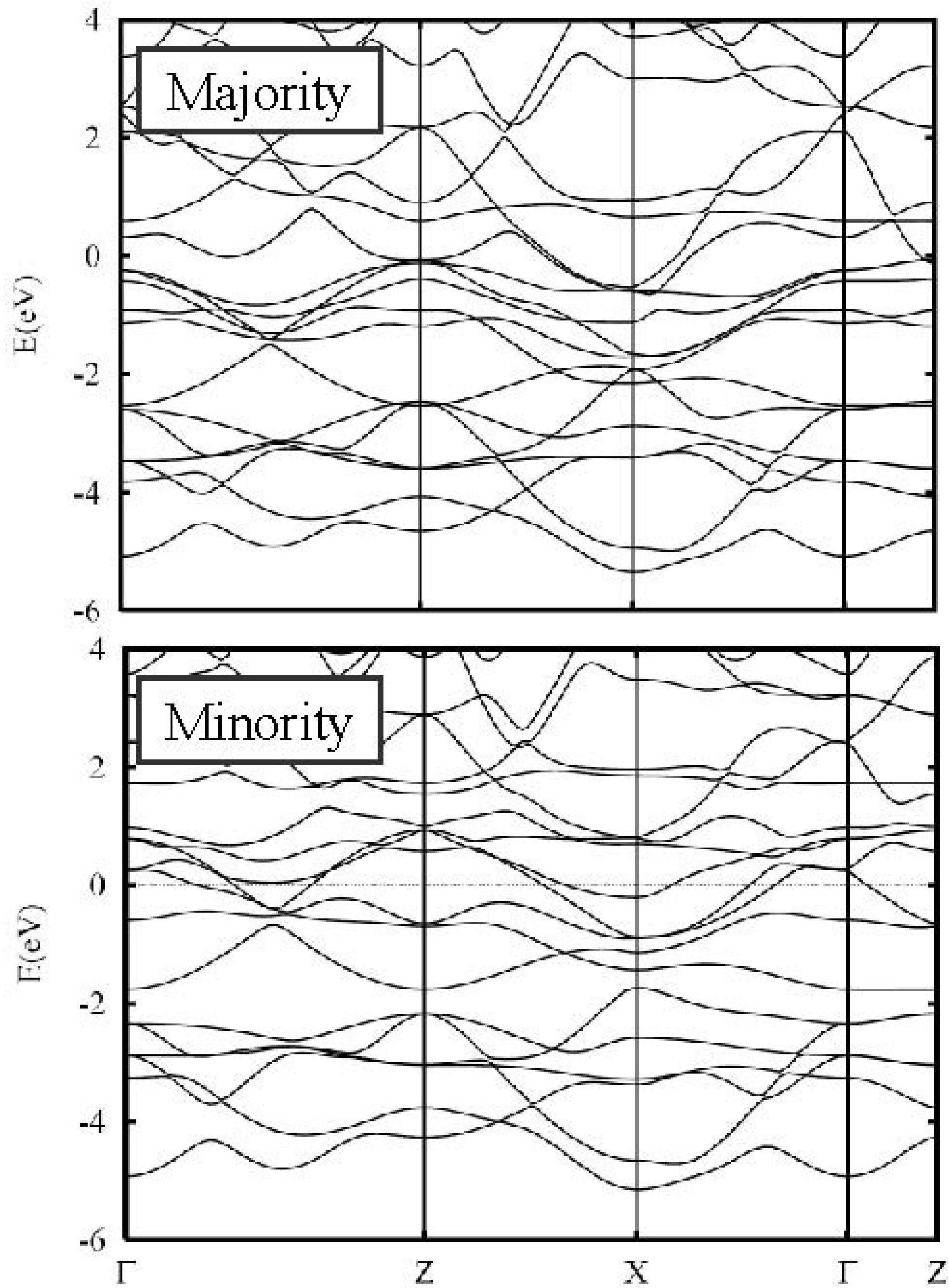

Fig. 7: Calculated majority and minority spin band structures of checkerboard ordered ferromagnetic $BaFe_{2-x}Cr_xAs_2$ with x = 1. The directions shown are in the bct basal plane and along the $k_z$ direction, as described in Ref. [12].

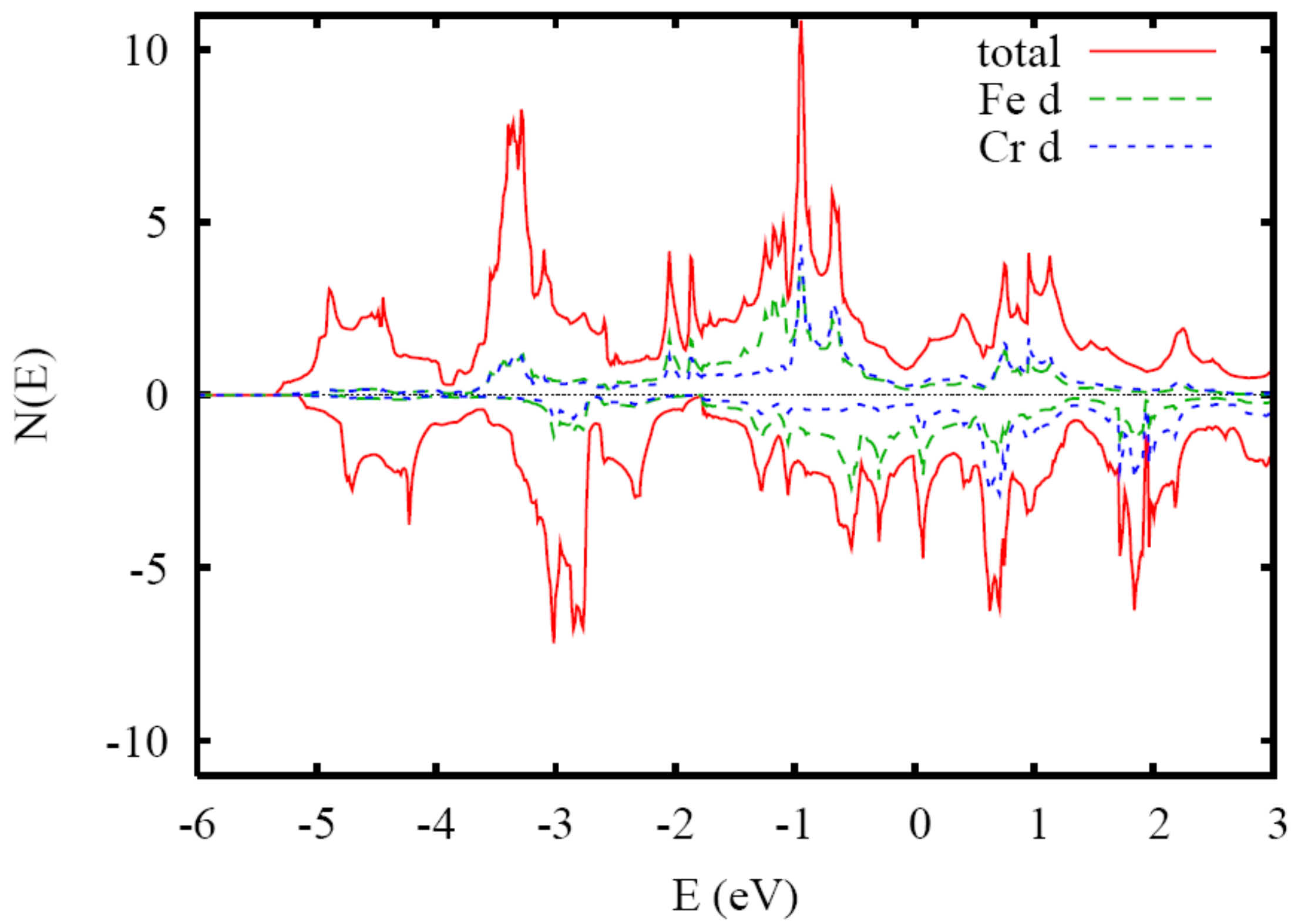

Fig. 8: (Color online) Calculated electronic DOS of ferromagnetic $BaFe_{2-x}Cr_xAs_2$ with x = 1, on a per formula unit basis. Majority spin is above the axis and minority spin below. The projections are on a per atom basis.

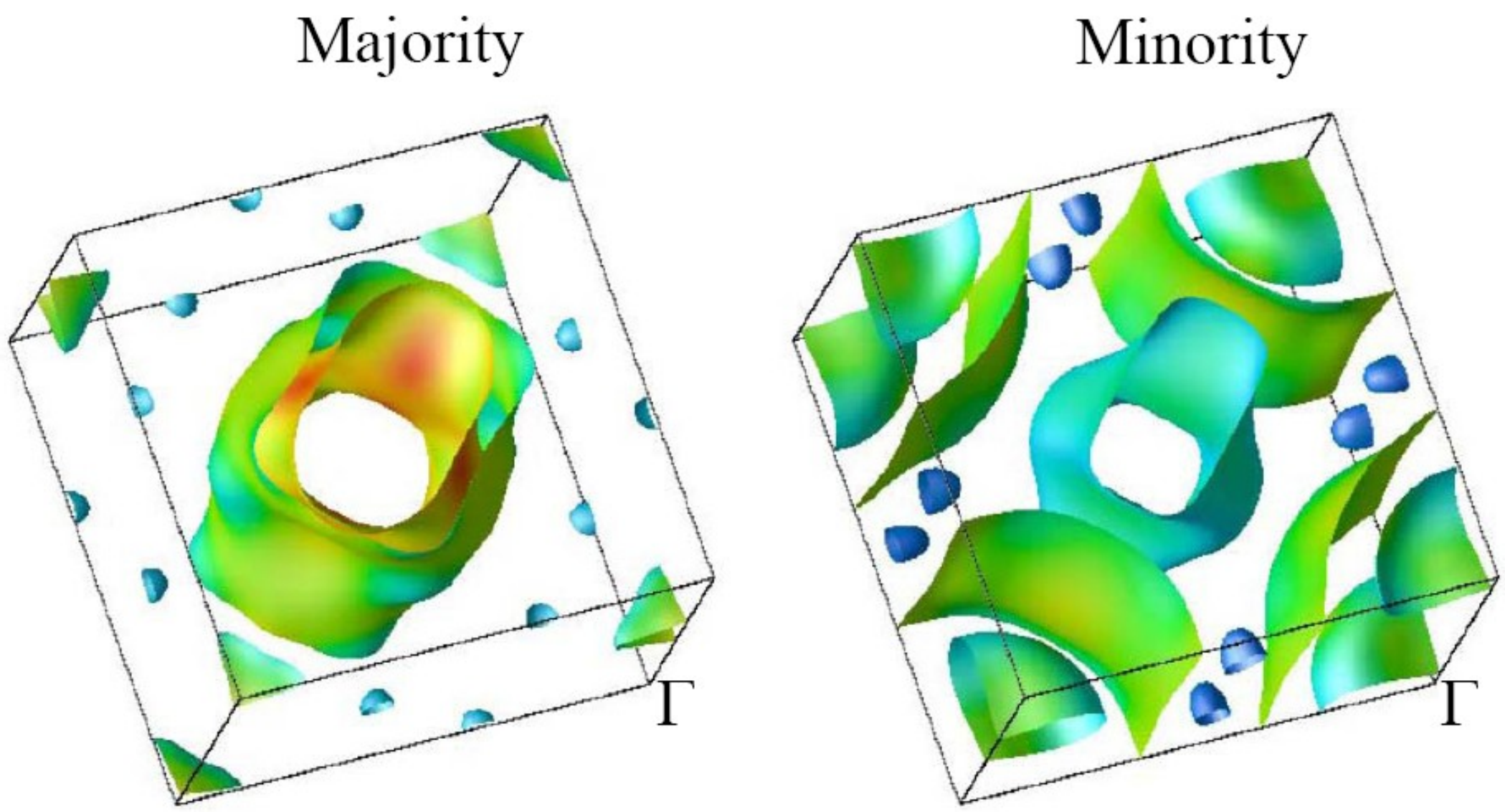

Fig. 9: (Color online) Calculated Fermi surface for majority (left) and minority (right) spin electrons in ordered ferromagnetic $BaFe_{2-x}Cr_xAs_2$ with x = 1. The shading is by band velocity; light blue represents low velocity bands and darker green represents high velocity bands.

Table 1: Chemical composition and lattice constants of $BaFe_{2-x}Cr_xAs_2$ found at room temperature. The lattice constants are determined from powder x-ray diffraction analyses.

| Nominal Fe:Cr | Microprobe Fe:Cr x (± 0.02) | *c* (Å) | *a* (Å) |
|---|---|---|---|
| 2:0 | 2:0 | 13.018(1) | 3.9642(3) |
| 1.94:0.06 | 1.96:0.04 | 13.040(1) | 3.9620(3) |
| 1.88:0.12 | 1.92:0.08 | 13.064(2) | 3.9631(4) |
| 1.78:0.22 | 1.86:0.14 | 13.085(2) | 3.9618(4) |
| 1.72:0.28 | 1.77:0.20 | 13.112(1) | 3.9665(3) |
| 1.68:0.32 | 1.77:0.23 | 13.120(2) | 3.9601(3) |
| 1.5:0.5 | 1.64:0.36 | 13.165(2) | 3.9657(3) |
| 1:1 | 1.25:0.75 | 13.260(1) | 3.9792(3) |
| 0:2 | 0:2 | 13.632(3) | 3.9678(5) |

Table 2: Crystal and structure data parameters for x = 0.04 in $BaFe_{2-x}Cr_xAs_2$ ($\lambda$ = 0.71073 Å).

| Temp. (K) | 293(2) | 173(2) | 119(2) | 117(2) | 100(2) |
|---|---|---|---|---|---|
| Crystal system | | | Tetragonal | | |
| Space group (no.), *Z* | | | *I*4/*mmm* (139), 2 | | |
| *a* (Å) | 3.970(2) | 3.964(2) | 3.967(2) | 3.962(2) | 3.954(2) |
| *c* (Å) | 13.022(10) | 13.002(11) | 12.999(11) | 12.984(10) | 12.963(10) |
| $V$ (Å$^3$) | 205.2(2) | 204.3(2) | 204.6(2) | 203.84(19) | 202.7(2) |
| $d_{calc}$. (Mg/m$^3$) | 6.454 | 6.483 | 6.474 | 6.499 | 6.536 |
| Final $R_1$, $wR_2$ [$I > 2\sigma(I)$] | 0.0295, 0.0565 | 0.0226, 0.0421 | 0.0263, 0.0459 | 0.0273, 0.0504 | 0.0281, 0.0468 |
| Largest diff. peak & hole | 1.65, -0.83 | 1.90, -1.05 | 2.32, -1.64 | 2.11, -1.13 | 2.78, -1.29 |